\documentclass{optica-article}

\journal{opticajournal} 

\articletype{Research Article}

\usepackage{lineno}

\begin{document}

\title{A Unified Perspective on Causality and One-Sided System Responses in Time and Space Across Physical and Fourier Domains}

\author{Mohammadreza Salehi,\authormark{1} Francesco Monticone\authormark{1}}

\address{\authormark{1}School of Electrical and Computer Engineering, Cornell University, Ithaca, NY 14853, USA}

\email{\authormark{*}francesco.monticone@cornell.edu} 


\begin{abstract*} 
The principle of causality has long been mathematically associated with the frameworks of Titchmarsh’s theorem and the Kramers--Kronig relations. While these relations arise naturally in the context of temporal system responses---ensuring that the effect of an applied field or force does not precede its cause---they have recently been shown to provide a pathway for realizing one-sided system responses in a variety of physical settings. In particular, one-sided frequency responses and one-sided wavevector responses have been successfully studied and engineered, enabling the prospect of numerous applications based on the complete suppression of backward scattering. In this work, we present a brief review of causality and its connection to these Fourier-domain analogs. We then turn our attention to the only remaining setting in which a one-sided system response may be explored: one-sided spatial nonlocality. We specifically investigate the possibility of realizing a one-sided spatial response within the widely used framework of nonlocal flat optics, where we uncover fundamental obstacles that hinder the achievement of such functionality in these structures. This, in turn, raises an intriguing open question: is one-sided spatial nonlocal response merely incompatible with the specific platform of nonlocal flat optics, or is it fundamentally forbidden by nature itself?

\end{abstract*}

\section{Introduction}
Materials and systems, regardless of their origin or physical context, are governed by the universal principle of causality \cite{Ref1, Ref2, Ref3}, whereby the response at any given time depends only on inputs applied at preceding times. In electromagnetics and optics, causality imposes fundamental constraints on constitutive parameters---including permittivity and permeability---thereby dictating how physical phenomena such as frequency dispersion and resonance manifest \cite{Ref4}. These constraints are of particular relevance to metamaterials and metasurfaces, whose engineered subwavelength structures are designed to produce tailored electromagnetic properties, yet remain inherently bounded by causality limits on the achievable response \cite{Ref19}. Specifically, causality establishes intrinsic relationships between the real and imaginary parts of a system's frequency response through the Kramers-Kronig relations \cite{Ref6, Ref7, Ref8}. By interconnecting fundamental properties such as dispersion, absorption, and bandwidth, these relations set the ultimate performance bounds for a wide range of electromagnetic devices and platforms \cite{Ref8-1}, including absorbers \cite{Ref9, Ref10, Ref11}, invisibility cloaks \cite{Ref12, Ref13, Ref14, Ref15}, near-field radiative heat transfer devices \cite{Ref16}, and temporally modulated media \cite{Ref17}. 

\begin{figure}

\includegraphics[width=\textwidth]{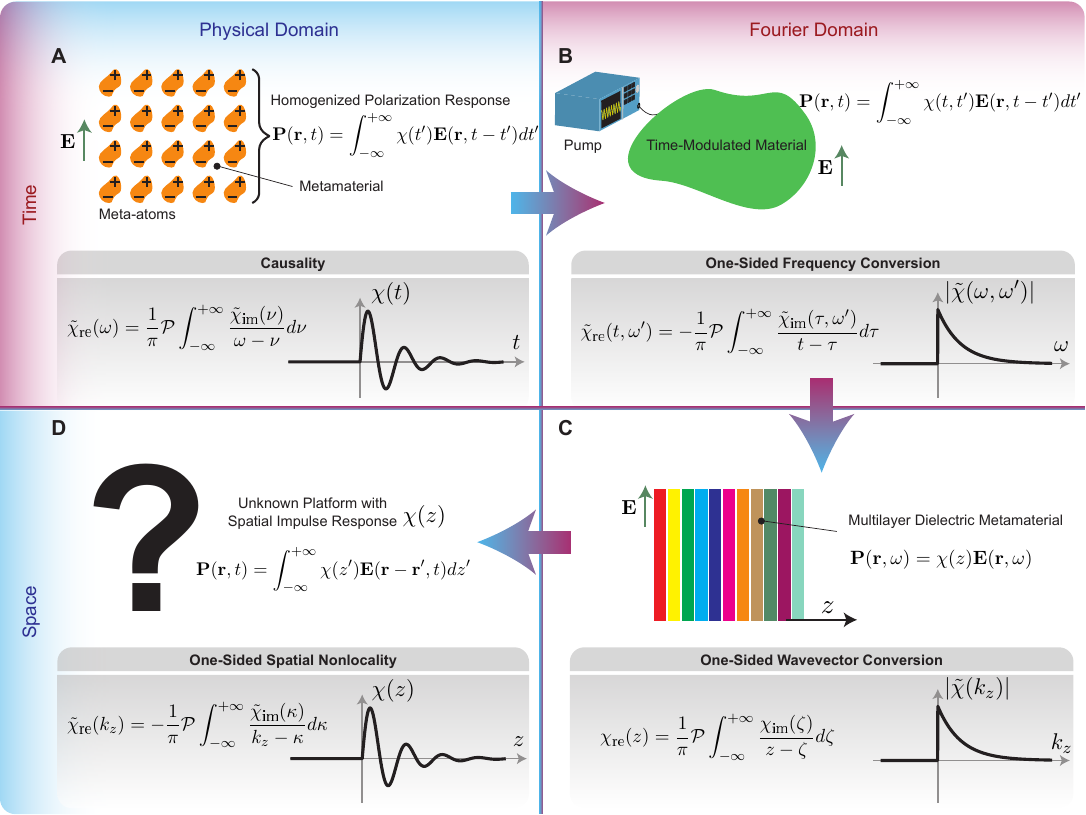}

\caption{Causality and its possible analogues organized into four quadrants spanning time and space along one dimension, and the physical and Fourier domains along the other. (A) The temporal impulse response of a metamaterial universally satisfies the principle of causality, which is mathematically equivalent to the existence of the Kramers--Kronig relations. (B) The Fourier-domain analogue of causality, referred to as spectral causality. A time-varying structure satisfying spectral causality imparts one-sided frequency conversion (e.g., exclusively up-conversion) on an input field. (C) The spatial analogue of spectral causality, realized in spatially varying structures satisfying the spatial Kramers--Kronig relations. This property enables wavevector conversion exclusively in one direction. (D) The remaining unexplored quadrant corresponds to one-sided spatial nonlocality. Whether and how such a response can be realized remains an open problem.}
\label{fig1}
\end{figure}

To gain insight into the mathematical framework underlying causality and its implications for physical systems and electromagnetic structures, we review the formalism established by Titchmarsh’s theorem. Consider a metamaterial (Fig.~\ref{fig1}A) whose homogenized electromagnetic response is characterized by the electric susceptibility $\chi(t)$, such that the effective macroscopic polarization is related to the applied electric field through the convolution relation $ \mathbf{P}(\mathbf{r},t)=\int_{-\infty}^{+\infty}\chi(t')\mathbf{E}(\mathbf{r},t-t')dt'$. Defining the Fourier transform according to $\tilde{\chi}(\omega)=\int_{-\infty}^{+\infty}\chi(t)\exp(-i\omega t) dt$, Titchmarsh's theorem states that \cite{Ref18, Ref19, Ref20} 

\textbf{Titchmarsh's Theorem.}
\textit{
If $\tilde{\chi}(\omega)=\tilde{\chi}_{\text{re}}(\omega)+i\tilde{\chi}_{\text{im}}(\omega)$ is square integrable and fulfills any one of the four conditions below, then it fulfills all the other conditions as well:
}
\begin{itemize}

    \item \textit{The inverse Fourier transform of $\tilde{\chi}(\omega)$ is causal:}
    \begin{equation} \label{eq1}
    \chi(t)=0, \qquad t<0 .
    \end{equation}

    \item \textit{
Replacing $\omega$ by $q\equiv x+iy$, the function $\tilde{\chi}(q)$ is analytic in the complex plane $q$ for $y<0$ and approaches $\tilde{\chi}(x)$ almost everywhere as $y\to0$. Furthermore, for all $y<0$
\begin{equation} \label{eq2}
\int_{-\infty}^{\infty} |\tilde{\chi}(x+iy)|^2 \, dx < k
\end{equation}
for some finite number $k$ (i.e., the integral is bounded).
}

    \item \textit{The first Kramers-Kronig relation applies:}
    \begin{equation}  \label{eq3}
    \tilde{\chi}_{\text{re}}(\omega)
    =
    \frac{1}{\pi}
    \mathcal{P}
    \int_{-\infty}^{\infty}
    \frac{\tilde{\chi}_{\text{im}}(\nu)}
    {\omega-\nu}
    \, d\nu .
    \end{equation}

    \item \textit{The second Kramers-Kronig relation applies:}
    \begin{equation} \label{eq4}
    \tilde{\chi}_{\text{im}}(\omega)
    =
    -\frac{1}{\pi}
    \mathcal{P}
    \int_{-\infty}^{\infty}
    \frac{\tilde{\chi}_{\text{re}}(\nu)}
    {\omega-\nu}
    \, d\nu .
    \end{equation}
    
\end{itemize}
\noindent The Kramers–Kronig relations intrinsically link the real and imaginary parts of $\tilde{\chi}(\omega)$, implying that the loss/gain characteristics of a metamaterial, represented by $\tilde{\chi}_{\text{im}}(\omega)$, uniquely determine its dispersive phase/refractive response, represented by $\tilde{\chi}_{\text{re}}(\omega)$. In other words, when engineering a specific loss/gain profile for a metamaterial, the corresponding dispersion cannot be chosen independently, as it is fundamentally constrained by the requirement of causality. Throughout this paper, applications of Titchmarsh's theorem assume the temporal Fourier transform convention $\tilde{\chi}(\omega)=\int_{-\infty}^{+\infty}\chi(t)\exp(-i\omega t)\,dt$ and the spatial convention $\tilde{\chi}(k_z)=\int_{-\infty}^{+\infty}\chi(z)\exp(ik_z z)\,dz$. When discussing one-sided responses in the Fourier domain, we employ the corresponding inverse Fourier transform.

One-sided system responses may also arise in a variety of other contexts. While nature fundamentally enforces a one-sided temporal response through causality, there exist three other scenarios in which a one-sided response may be realized: one-sided frequency conversion (the Fourier-domain analog of causality), one-sided wavevector/momentum conversion (the spatial Fourier-domain analog of causality), and one-sided nonlocal spatial response. These cases, outlined in Fig.~\ref{fig1}, are not inherently guaranteed by nature and instead require careful and deliberate engineering. It has been shown that one-sided frequency conversion can be achieved in time-varying structures satisfying a temporal analog of the Kramers–Kronig relations, formulated in the time domain rather than the frequency domain, thereby enforcing \emph{spectral causality} \cite{Ref21}. Likewise, when a spatial analog of the Kramers–Kronig relations is employed in the design of spatially inhomogeneous structures, one-sided momentum conversion naturally emerges \cite{Ref22}. In this work, we review these approaches and discuss their implications for electromagnetics and optics from an application-oriented perspective. We then address the still-open problem of one-sided spatial nonlocality by investigating its feasibility within the well-established framework of nonlocal flat optics \cite{Ref33, Ref34}, examining the fundamental reasons why this platform fails to realize such a response, and ultimately raising the broader question of whether one-sided spatial nonlocality can be realized in nature at all. The answer may well lie in the rich field of metamaterials, which, since its inception, has provided particularly fertile ground for the exploration of such extreme and anomalous electromagnetic phenomena, with pioneering contribution by Prof. Engheta on artificial spatial nonlocality more than two decades ago \cite{Ref43, Ref44}. 

\section{Spectral Causality: One-Sided Frequency Conversion}

We begin by reviewing the concept of spectral causality introduced in \cite{Ref21}. Consider a time-varying, nonmagnetic material characterized by the general constitutive relation $\mathbf{D}(\mathbf{r},t)=\varepsilon_0[\varepsilon_s \mathbf{E}(\mathbf{r},t)+\int_{-\infty}^{+\infty}\chi(t, t')\mathbf{E}(\mathbf{r},t-t')dt']$ as shown in Fig.~\ref{fig1}B, where $\chi(t,t')$ denotes the time-dispersive, time-varying electric susceptibility, $\varepsilon_s$ is the static and spatially homogeneous background relative permittivity, and $\varepsilon_0$ is the permittivity of free space \cite{Ref23, Ref24}. The vector wave equation governing the electric field is given by
\begin{equation} \label{eq5}
\nabla \times \nabla \times \mathbf{E}+\mu_0\frac{\partial^2 \mathbf{D}}{\partial t^2}=0.
\end{equation}
For a transverse electromagnetic (TEM) wave with wavevector $\mathbf{k}$, the spatial differential operator reduces according to $\nabla \rightarrow -i\mathbf{k}$, yielding the following time-domain equation:
\begin{equation} \label{eq6}
k^2E(t)+\frac{1}{c_s^2}\frac{d^2}{dt^2}\left[E(t)+\frac{1}{\varepsilon_s}\int_{-\infty}^{+\infty}\chi(t,t')E(t-t')dt'\right]=0,
\end{equation}
where $k=||\mathbf{k}||$ and $c_s=1/\sqrt{\mu_0\varepsilon_0\varepsilon_s}$ denotes the speed of light in the material in the absence of modulation, i.e., when $\chi(t,t')=0$. Viewing the modulation as a temporal perturbation, the problem can be formulated within a scattering framework. In the absence of modulation, the solution to Eq.~(\ref{eq6}) is $E_{\text{inc}}(t)=E_0 e^{ikc_s t}$; therefore, the full field can be expressed as a superposition of the incident and scattered contributions:
\begin{equation} \label{eq7}
E(t)=E_{\text{inc}}(t)+E_{\text{sca}}(t).
\end{equation}
Substituting this expression into Eq.~(\ref{eq6}) and multiplying both sides by $c_s^2$, we obtain
\begin{equation} \label{eq8}
k^2c_s^2 E_{\text{sca}}(t)+\frac{d^2 E_{\text{sca}}(t)}{dt^2}
=-\frac{1}{\varepsilon_s}\frac{d^2}{dt^2}\int_{-\infty}^{+\infty}\chi(t,t') E(t-t') dt'.
\end{equation}
To understand the implications of this equation in terms of frequency conversion, we take its Fourier transform. For the convolution-like term on the right-hand side, we obtain
\begin{equation} \label{eq9}
\int_{-\infty}^{+\infty}\left[\int_{-\infty}^{+\infty}\chi(t,t')E(t-t')dt'\right]e^{-i\omega t} dt
=\int_{-\infty}^{+\infty}\left[\int_{-\infty}^{+\infty}\chi(t,t')E(t-t')e^{-i\omega t} dt\right]dt'.
\end{equation}
Interchanging the order of integration is justified provided that $\chi(t,t')$ is absolutely integrable, in the sense that (Fubini’s theorem \cite{Ref25, Ref26})
\begin{equation} \label{eq10}
\int_{-\infty}^{+\infty}\int_{-\infty}^{+\infty}|\chi(t,t')| dt dt' < \infty.
\end{equation}
Under this condition, Eq.~(\ref{eq9}) can be further simplified using the Fourier transform multiplication (convolution) property, yielding
\begin{equation} \label{eq11}
\int_{-\infty}^{+\infty}\chi(t,t')E(t-t')e^{-i\omega t} dt
=\frac{1}{2\pi}\int_{-\infty}^{+\infty}\tilde{\chi}(\omega-\omega',t') \tilde{E}(\omega')e^{-i\omega' t'} d\omega'.
\end{equation}
Equation~(\ref{eq8}) can therefore be expressed in the Fourier domain as
\begin{equation} \label{eq12}
\begin{aligned}
\tilde{E}_{\text{sca}}(\omega)
&=\frac{\omega^2}{2\pi\varepsilon_s\left(k^2c_s^2-\omega^2\right)}
\int_{-\infty}^{+\infty}\int_{-\infty}^{+\infty}
\tilde{\chi}(\omega-\omega',t')\,\tilde{E}(\omega')e^{-i\omega' t'}\,d\omega'\,dt' \\
&=\frac{\omega^2}{2\pi\varepsilon_s\left(k^2c_s^2-\omega^2\right)}
\int_{-\infty}^{+\infty}
\tilde{\chi}(\omega-\omega',\omega')\,\tilde{E}(\omega')\,d\omega',
\end{aligned}
\end{equation}
where $\tilde{\chi}(\omega,\omega')$ denotes the Fourier transform of $\chi(t,t')$ with respect to both variables $t$ and $t'$. This can alternatively be expressed by introducing the linear operator $\mathbf{U}$:
\begin{equation} \label{eq13}
\tilde{E}_{\text{sca}}(\omega)=\mathbf{U}\tilde{E}(\omega)=\mathbf{U}\big[\tilde{E}_{\text{inc}}(\omega)+\tilde{E}_{\text{sca}}(\omega)\big],
\end{equation}
which can be rearranged and solved for $\tilde{E}_{\text{sca}}(\omega)$ via operator inversion as
\begin{equation} \label{eq14}
(\mathbf{I}-\mathbf{U})\tilde{E}_{\text{sca}}(\omega)=\mathbf{U}\tilde{E}_{\text{inc}}(\omega)
\quad \Rightarrow \quad
\tilde{E}_{\text{sca}}(\omega)=(\mathbf{I}-\mathbf{U})^{-1}\mathbf{U}\tilde{E}_{\text{inc}}(\omega).
\end{equation}
Using the Neumann series expansion \cite{Ref27, Ref28}
\begin{equation} \label{eq15}
(\mathbf{I}-\mathbf{U})^{-1}=\mathbf{I}+\mathbf{U}+\mathbf{U}^2+\cdots, \qquad ||\mathbf{U}||<1,
\end{equation}
the solution can then be expressed as
\begin{equation} \label{eq16}
\tilde{E}_{\text{sca}}(\omega)=\mathbf{U}\tilde{E}_{\text{inc}}(\omega)+\mathbf{U}^2 \tilde{E}_{\text{inc}}(\omega)+\mathbf{U}^3 \tilde{E}_{\text{inc}}(\omega)+\cdots.
\end{equation}

This formalism has the advantage of expressing the scattered field through successive applications of a single linear operator to the incident field. Consequently, it is sufficient to examine only the first few terms of the series in order to capture the essential dynamics governing frequency conversion in the structure. This approach, commonly referred to as the Born approximation \cite{Ref29, Ref30, Ref31, Ref32}, is widely used and highly accurate in regimes where the modulation---whether temporal or spatial---constitutes a weak perturbation. According to Eq.~(\ref{eq12}), the linear operator $\mathbf{U}$ involves a convolution in the frequency domain. The scattered field at a given frequency $\omega$ depends on $\tilde{E}(\omega')$ over the entire frequency spectrum, weighted by the kernel $\tilde{\chi}(\omega-\omega',\omega')$. One-sided frequency up-conversion can be realized if the kernel $\tilde{\chi}(\omega-\omega',\omega')$ vanishes for $\omega<\omega'$. Equivalently, the Fourier-domain system function must be \emph{spectrally causal} in its first argument:
\begin{equation} \label{eq17}
\tilde{\chi}(\omega,\omega')=0, \qquad \text{for\ } \omega<0 \text{\ for all\ } \omega'.
\end{equation}
This condition is mathematically analogous to temporal causality; therefore, provided square integrability is satisfied, Titchmarsh’s theorem can be applied to relate spectral causality in $\omega$ to Kramers--Kronig relations in $t$. We employ the same mathematical framework as before, but with a slightly modified Fourier-transform convention, $\tilde{\chi}(t,\omega')=\frac{1}{2\pi}\int_{-\infty}^{+\infty}\tilde{\chi}(\omega,\omega')\exp(i\omega t) dt$, which corresponds to the inverse Fourier transform in the standard formulation of Titchmarsh’s theorem. This change in sign convention leads to corresponding modifications in the theorem’s implications, including analyticity in the upper half-plane (rather than the lower half-plane) and a sign reversal in the Kramers--Kronig relations. For example,
\begin{equation} \label{eq18}
 \tilde{\chi}_{\text{re}}(t, \omega')=-\dfrac{1}{\pi} \mathcal{P}\int_{-\infty}^{+\infty} \dfrac{\tilde{\chi}_{\text{im}}(\tau, \omega')}{t-\tau}d\tau, \quad \text{\ for all\ } \omega'.
\end{equation}
Equation~(\ref{eq18}) therefore provides a pathway for engineering time-varying structures that enable frequency conversion exclusively in one direction. It is important to note, however, that any engineered material response $\chi(t,t')$ must also satisfy the universal requirement of temporal causality. In particular, the convolution in Eq.~(\ref{eq8}) must remain causal, which requires that $\chi(t,t')=0$ for $t'<0$. Consequently, the material response must satisfy the conventional Kramers--Kronig relations with respect to the frequency variable $\omega'$.

\begin{figure}

\includegraphics[width=\textwidth]{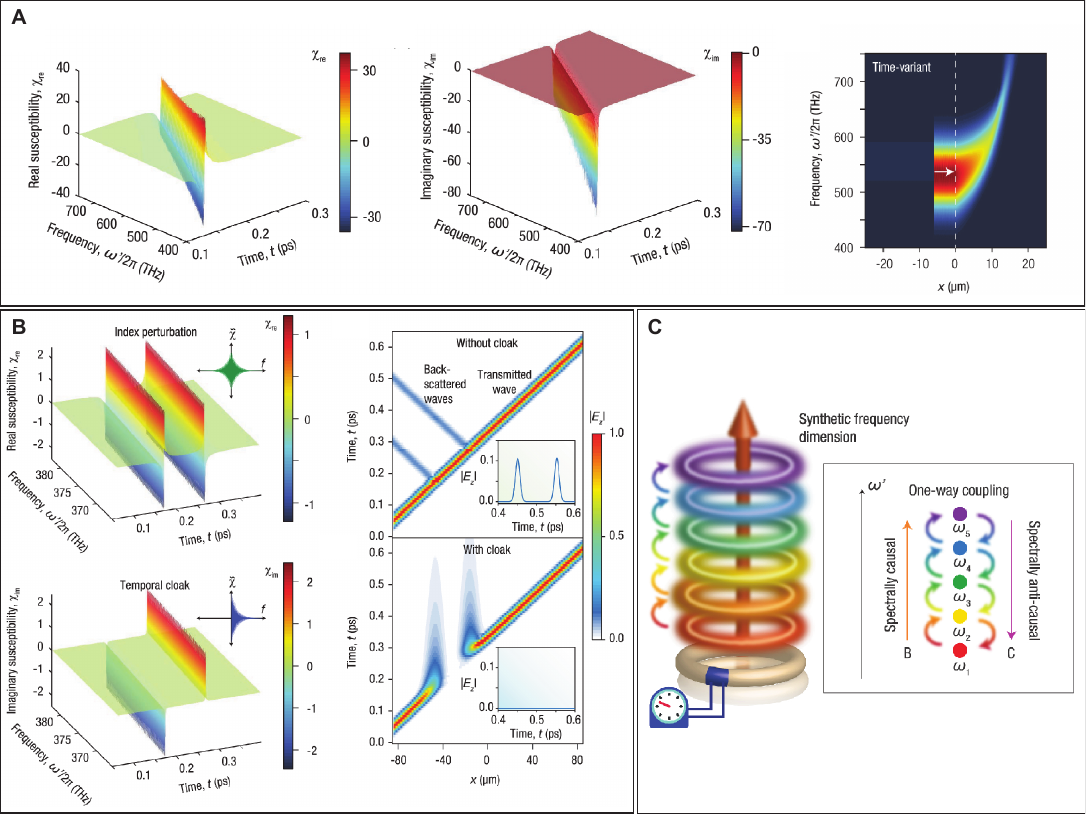}

\caption{Various applications of the concept of spectral causality. (A) Demonstration of broadband reflectionless absorption achieved through a spectrally causal modulation of a lossy material according to Eq.~(\ref{eq19}). The FDTD result, shown at a particular instant in time, confirms the absence of reflection, since one-sided frequency up-conversion prevents the generation of negative-frequency components and therefore suppresses back-scattering. (B) Demonstration of temporal cloaking of a refractive-index perturbation through the introduction of a spectrally causal gain--loss modulation based on Eq.~(\ref{eq20}). This suppresses scattering induced by the temporal perturbation, thereby effectively cloaking the perturbation event. (C) Spectrally causal modulation of a resonator enables unidirectional transport along a synthetic frequency dimension. Without requiring spatiotemporal traveling-wave modulations, coupling between higher- and lower-frequency modes can be selectively suppressed in either direction. Panels (A)–(C) are reproduced with permission from \cite{Ref21}, Optica.}
\label{fig2}
\end{figure}

One application of one-sided frequency conversion is in the realization of broadband reflectionless absorbers (Fig.~\ref{fig2}A). By engineering a temporally modulated lossy susceptibility that satisfies Eq.~(\ref{eq18}) while also respecting temporal causality, the system is prevented from generating lower-frequency or negative-frequency scattered components. As a result, an incident wave $E_{\text{inc}}(t)=E_0 e^{ikc_s t}$ undergoes continuous frequency up-conversion without back-scattering (negative-frequency components), leading to near-perfect broadband absorption. It can be shown that such a device can indeed be implemented using a susceptibility of the form \cite{Ref21}
\begin{equation} \label{eq19}
\tilde{\chi}(t,\omega') =
\frac{\omega_p^2}
{\left(\omega_0 - K (t - t_{\mathrm{offset}})\right)^2 - \omega'^2 + i \gamma \omega' },
\end{equation}
which corresponds to a standard Lorentz-type dispersion characterized by plasma frequency $\omega_p$, damping rate $\gamma$, and a time-varying resonance frequency $\omega_0'(t)=\left|\omega_0 - K (t - t_{\text{offset}})\right|$. This Lorentz-type response has the advantage of being temporally causal, since it satisfies the conventional Kramers--Kronig relations with respect to $\omega'$. Moreover, for all values of $\omega'$, the susceptibility maintains an approximately Lorentzian dependence on $t$ within a certain time window determined by $t_{\text{offset}}$, and therefore approximately satisfies Eq.~(\ref{eq18}) as well.

A second application that may be envisioned for spectral causality is temporal cloaking \cite{Ref21}, in which gain--loss modulations satisfying Eq.~(\ref{eq18}) are used to suppress the scattering induced by a temporal refractive-index perturbation (Fig.~\ref{fig2}B). When a medium undergoes a temporal refractive-index perturbation represented by $\tilde{\chi}_{\text{re}}(t,\omega')$, a wave propagating through the medium experiences backward scattering, thereby revealing the presence of the perturbation to an observer located at the source of the incident wave. This effect can be mitigated by coordinating the refractive-index perturbation with a properly engineered gain--loss modulation determined through the second Kramers--Kronig relation for spectral causality:
\begin{equation} \label{eq20}
\tilde{\chi}_{\text{im}}(t, \omega')=\dfrac{1}{\pi}\mathcal{P}\int_{-\infty}^{+\infty} \dfrac{\tilde{\chi}_{\text{re}}(\tau, \omega')}{t-\tau} d\tau.
\end{equation}
The introduction of this tailored gain--loss modulation suppresses the backward-scattered field at the incident frequency (Fig.~\ref{fig2}B), effectively cloaking the refractive-index perturbation from an observer at the origin of the incident wave.

Reference \cite{Ref21} further demonstrates that spectral causality can enable unidirectional transport along synthetic frequency dimensions in dynamically modulated resonators (Fig.~\ref{fig2}C). By employing spatially localized temporal modulations satisfying spectral causality or spectral anti-causality, coupling between resonator modes can be restricted to exclusively blue- or red-shifting transitions, respectively. Unlike conventional approaches based on periodic phase modulation and resonance detuning \cite{Ref32-1}, this mechanism arises from the interplay between refractive-index and loss modulations, enabling flexible control of frequency transport. 

\section{One-Sided Momentum Conversion}

\begin{figure}

\includegraphics[width=\textwidth]{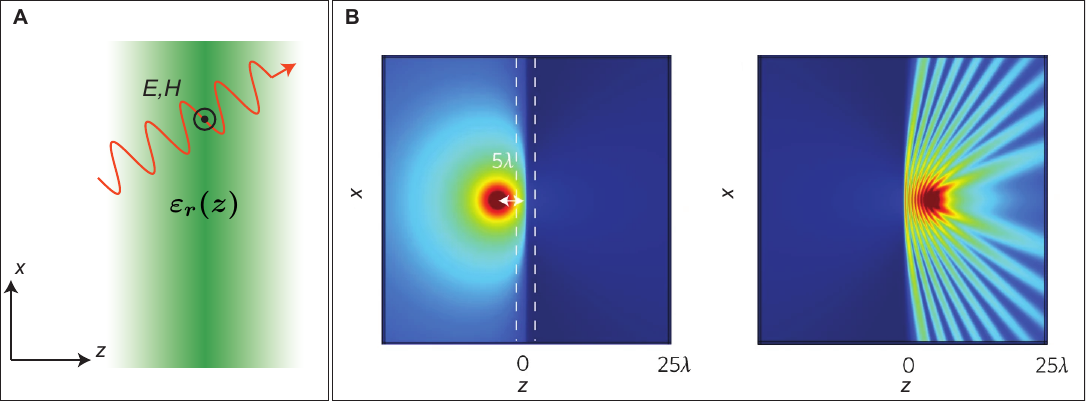}

\caption{One-sided wavevector conversion in a spatially varying structure engineered through the spatial Kramers--Kronig relations. (A) Schematic of a reflectionless absorber designed to absorb all waves propagating in the positive $z$ direction, irrespective of their propagation angle with respect to the $x$ axis. (B) COMSOL simulations confirm that all incident waves originating from a two-dimensional point source placed to the left of the absorber are absorbed without any back-scattering. Since the structure supports only wavevector up-conversion (i.e., scattering toward positive $z$), a source placed to the right of the absorber experiences substantial reflection. Panel (B) is reproduced with permission from \cite{Ref22}, Springer Nature.}
\label{fig3}
\end{figure}

The principles of spectral causality can also be extended to spatially varying structures, analogous to their time-varying counterparts \cite{Ref22}. Consider a time-harmonic problem with the convention $e^{i\omega t}$, in which the medium is characterized by a spatially varying electric susceptibility $\chi(z)$ embedded in a background medium with relative permittivity $\varepsilon_b$ (Fig.~\ref{fig3}A). The corresponding constitutive relation is given by $\mathbf{D}(\mathbf{r},\omega)=\varepsilon_0\varepsilon_r(z)\mathbf{E}(\mathbf{r},\omega)=\varepsilon_0[\varepsilon_b \mathbf{E}(\mathbf{r},\omega)+\chi(z)\mathbf{E}(\mathbf{r},\omega)]$. For transverse electric (TE) waves propagating at an arbitrary angle, with electric field profile of the form $\mathbf{E}(\mathbf{r},\omega)=E_y(z,\omega)e^{-ik_x x}\hat{\mathbf{y}}$, Maxwell’s equations reduce to
\begin{equation} \label{eq21}
\frac{d^2 E_y(z)}{dz^2}+\left[k_0^2\varepsilon_b-k_x^2+k_0^2\chi(z)\right]E_y(z)=0,
\end{equation}
where $k_0^2=\omega^2\mu_0\varepsilon_0$. Again adopting a scattering formulation, the total field may be expressed as
\begin{equation} \label{eq22}
E_y(z)=E_0 e^{-i\sqrt{k_0^2\varepsilon_b-k_x^2}\,z}+E_{y,\mathrm{sca}}(z),
\end{equation}
where $E_{y,\mathrm{sca}}(z)$ denotes the scattered field. Substituting Eq.~(\ref{eq22}) into Eq.~(\ref{eq21}), and transforming to the Fourier domain using the convention $\tilde{E}_y(k_z)=\int_{-\infty}^{+\infty} E_y(z) e^{ik_zz}dz$, yields
\begin{equation} \label{eq23}
\tilde{E}_{y,\mathrm{sca}}(k_z)
=\frac{k_0^2}{2\pi\left(k_0^2\varepsilon_b-k_x^2-k_z^2\right)}
\int_{-\infty}^{+\infty}
\tilde{\chi}(k_z-k_z')
\tilde{E}_y(k_z')\,dk_z'.
\end{equation}
This result is directly analogous to Eq.~(\ref{eq12}) obtained for the time-varying case, and therefore the same reasoning applies here. In particular, the scattered field contains only wavenumbers larger than that of the incident wave provided that
\begin{equation} \label{eq24}
\tilde{\chi}(k_z)=0, \qquad \text{for } k_z<0.
\end{equation}
Invoking Titchmarsh’s theorem, and taking into account the chosen sign convention for the spatial Fourier transform, Eq.~(\ref{eq24}) implies the following spatial Kramers--Kronig relation
\begin{equation} \label{eq25}
\chi_{\mathrm{re}}(z)=
\frac{1}{\pi}\mathcal{P}\int_{-\infty}^{+\infty}
\frac{\chi_{\mathrm{im}}(\zeta)}{z-\zeta}\,d\zeta.
\end{equation}
Consequently, spatial material profiles engineered to satisfy Eq.~(\ref{eq25}) exhibit suppression of back-scattering for TE-polarized waves propagating in the positive $z$ direction at arbitrary angles with respect to the $x$ axis. This property can be exploited to realize an angle-independent reflectionless slab that functions as an efficient absorber by prescribing a sufficiently large loss profile $\chi_{\mathrm{im}}(z)$ and a sufficiently large slab thickness, while determining the corresponding real part from Eq.~(\ref{eq25}). One such susceptibility profile is given by \cite{Ref22}
\begin{equation} \label{eq26}
\chi(z)=-A\frac{i+z/\xi}{1+(z/\xi)^2},
\end{equation}
where $\xi$ determines the characteristic spatial scale of the profile, and $A$ specifies its amplitude. As shown in Fig.~\ref{fig3}B, when a point source is placed to the left of the lossy slab, all waves incident on the structure are absorbed without reflection. In contrast, when the source is positioned to the right of the slab, scattering is no longer suppressed, as evidenced by the nonuniform radiation pattern observed in the right-hand region. The resulting reflectionless absorber is therefore inherently unidirectional (with respect to its reflection response, not its transmission response, which would require nonreciprocity), although the preferred direction can be reversed by introducing a negative sign in the spatial Kramers--Kronig relation.

While the spatial Kramers--Kronig relation in Eq.~(\ref{eq25}) guarantees reflectionless operation for TE polarization, this property does not directly extend to transverse magnetic (TM) polarization. To understand this distinction, we note that the TM field satisfies \cite{Ref22}
\begin{equation} \label{eq27}
\nabla \cdot \left[\varepsilon_r^{-1}(z)\nabla H_y\right] + k_0^2 H_y = 0,
\end{equation}
where $\varepsilon_r^{-1}(z)$ denotes the reciprocal permittivity profile. Spatial engineering based on Kramers--Kronig relations enforces analyticity of $\chi(z)$, and consequently of $\varepsilon_r(z)$, in the lower half of the complex $z$-plane via Titchmarsh’s theorem. However, in the TM case, Eq.~(\ref{eq27}) shows that the relevant quantity is instead $\varepsilon_r^{-1}(z)$. Therefore, achieving back-scattering suppression for both polarizations requires that both $\varepsilon_r(z)$ and $\varepsilon_r^{-1}(z)$ be analytic in the lower half-plane. This condition implies that $\varepsilon_r(z)$ must be free of zeros and poles in the lower half of the complex $z$-plane. Such a permittivity profile can be constructed by using the following loss distribution, which can then be substituted into Eq.~(\ref{eq25}) to obtain the corresponding full permittivity function \cite{Ref22}:
\begin{equation} \label{eq28}
\operatorname{Im}[\varepsilon_r(z)] =-
\frac{h(L - z)}{4L}
\left[1 + \operatorname{erf}\left(\frac{z}{\xi}\right)\right]
\left[1 + \operatorname{erf}\left(\frac{L - z}{\xi}\right)\right].
\end{equation}
The parameters $h$, $L$, and $\xi$ control the amplitude, spatial extent, and smoothness of the loss profile, respectively.

In addition to enabling the implementation of one-sided momentum conversion, the spatial Kramers--Kronig relations have also proven useful in addressing the phase problem in scattering and diffraction measurements \cite{Ref49, Ref50, Ref51, Ref52}. In this problem, the phase of a complex-valued wavefront, $f(z)=|f(z)|e^{i\phi(z)}$, is reconstructed from intensity measurements, $|f(z)|$. The wavefront $f(z)$ is assumed to be bandlimited (i.e., to have compact support in wavevector space), which, by the Paley–Wiener theorem, implies that $f(z)$ is analytic throughout the entire complex $z$-plane. This assumption is well justified in far-field measurements, where spectral components with $|k_z|>k_0$ correspond to evanescent waves that decay rapidly and are therefore negligible. The amplitude and phase of $f(z)$ can be naturally combined into the real and imaginary parts of a single complex function by taking its logarithm, $\log f(z)=\log |f(z)|+i\phi(z)$. If $f(z)$ is guaranteed to be free of zeros in the lower half of the complex $z$-plane, then the spatial Kramers--Kronig relations can be applied to $\log f(z)$, providing a direct means of retrieving the phase $\phi(z)$ from intensity measurements:
\begin{equation} \label{eq28-1}
\phi(z)=-
\frac{1}{\pi}\mathcal{P}\int_{-\infty}^{+\infty}
\frac{\log |f(\zeta)|}{z-\zeta}\,d\zeta.
\end{equation}
In practice, however, solving the phase problem is considerably more challenging, since the assumption that $f(z)$ is free of zeros in the lower half-plane is generally not satisfied. Zeros in the lower half-plane introduce logarithmic singularities in $\log f(z)$, and the associated residue contributions must be accounted for to correct the phase obtained from Eq.~(\ref{eq28-1}) \cite{Ref51}. More sophisticated methods have therefore been developed to circumvent this difficulty. Examples include the Gerchberg--Saxton algorithm \cite{Ref53} and approaches based on Rouche's theorem, in which a suitably chosen known function is added to $f(z)$ to eliminate the lower half-plane zeros of the resulting superposition \cite{Ref51}.

\section{One-Sided Spatial Nonlocality}

Now that we have reviewed the three previously explored quadrants in the taxonomy of one-sided system responses, as illustrated in Fig.~\ref{fig1}, we draw the reader’s attention to the currently open problem of realizing one-sided spatial nonlocality. First, let us examine the problem in the context of material susceptibility, in order to preserve continuity with the preceding discussions. Consider a material with a spatially nonlocal response characterized by $\mathbf{D}(\mathbf{r},t)=\varepsilon_0[\varepsilon_b \mathbf{E}(\mathbf{r},t)+\int_{-\infty}^{+\infty}\chi(z-z')\mathbf{E}(\mathbf{r}',t)\, dz']$. In such a medium, the electric displacement at a given point in space depends not only on the local electric field, but also on the field values over an extended interval of $z$-positions \cite{Ref33}. In a manner analogous to the previous discussions, one may envision a one-sided nonlocal response, which would require the following Kramers--Kronig relation (with $\tilde{\chi}(k_z)=\int_{-\infty}^{+\infty} \chi(z)e^{ik_z z}dz$):
\begin{equation} \label{eq29}
\tilde{\chi}_{\text{re}}(k_z)=-\dfrac{1}{\pi}\mathcal{P}\int_{-\infty}^{+\infty} \dfrac{\tilde{\chi}_{\text{im}}(\kappa)}{k_z-\kappa}d\kappa.
\end{equation}
Unlike the Kramers--Kronig relations outlined in Sections 2 and 3, which provide a pathway for engineering material responses in either the temporal or spatial domains, the relation in Eq.~(\ref{eq29}) does not lend itself to a straightforward physical implementation. This is primarily because it requires direct control over the material response in the momentum domain, which is considerably less intuitive and less accessible than engineering in real space or time domain.

That said, several approaches to momentum-domain engineering have been explored in the context of inhomogeneous structures \cite{Ref47, Ref48, Ref34, Ref35}. Specifically, one of the most widely used platforms for realizing strong artificial nonlocality, namely, a strongly wavevector-dependent response, is multilayer slab structures, which constitute one of the foundational platforms of the relatively young field of nonlocal flat optics \cite{Ref34, Ref35}. When a wavefront impinges onto such a multilayer structure, the transmitted field on the opposite side forms a wavefront through angle-dependent phase/amplitude modulation imparted to the Fourier components of the incident field, i.e., its transverse-wavevector spectrum. The overall response is therefore equivalent to a spatial convolution with a nonlocal kernel. A natural question then arises: can Titchmarsh’s theorem and the associated Kramers--Kronig relations be exploited to engineer this wavevector-dependent response such that the resulting nonlocal interaction in real space becomes inherently one-sided? If possible, this would imply a nonlocal response in which the output at a given point depends on the input over an extended, but one-sided, spatial region, making the output effectively insensitive to a certain region of input space.

\begin{figure}

\includegraphics[width=\textwidth]{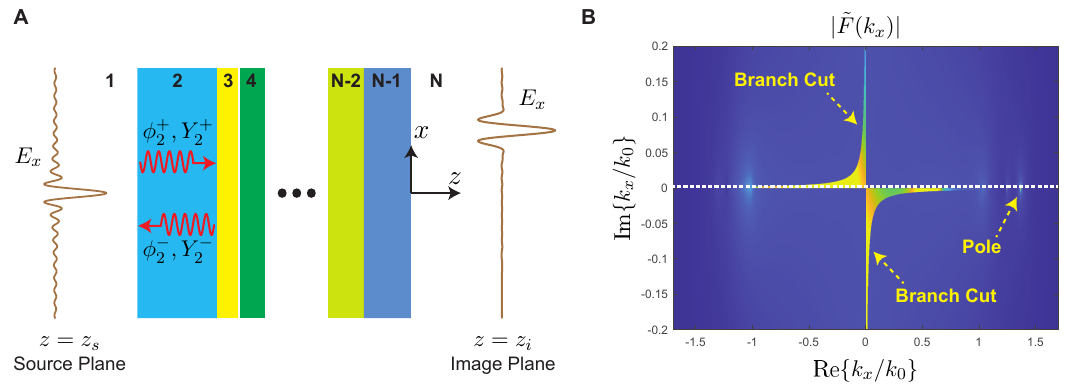}

\caption{Investigation of one-sided spatial nonlocality in multilayer structures widely employed within the framework of nonlocal flat optics. (A) Schematic of a hypothetical multilayer structure designed to produce the image of a planar source through convolution with a one-sided spatial response function $F(x)$. The spatial Fourier transform of the response, denoted by $\tilde{F}(k_x)$, is evaluated using a scattering-matrix formulation. The layers are anisotropic in a general form that does not permit polarization conversion (TE to TM or vice versa). The exterior regions are labeled $1$ and $N$, while the anisotropic (potentially nonreciprocal) layers are indexed from $2$ to $N-1$. (B) In the lossless case, poles and branch points exist on the real axis, while introducing infinitesimal loss into the exterior regions causes poles, branch points, and branch cuts to appear in $\tilde{F}(k_x)$ in both the upper and lower halves of the complex $k_x$ plane, thereby violating the conditions of Titchmarsh’s theorem in both the lossless and lossy cases. The numerical example considers $\varepsilon_1 = 1 - 0.005i$, $\varepsilon_N = 1.5$, and a single layer of thickness $L_2 = 1\,\mu\text{m}$ composed of a gyrotropic medium (e.g., a magnetized plasma) with $\varepsilon_{xx} = \varepsilon_{yy} = \varepsilon_{zz} = 4$ and $\varepsilon_{xz} = -\varepsilon_{zx} = -3i$, evaluated at the free-space wavelength $\lambda_0 = 1.55\,\mu\text{m}$.}
\label{fig4}
\end{figure}

Consider the geometry illustrated in Fig.~\ref{fig4}A, where dielectric layers are stacked along the $z$ direction and illuminated from the left by a source distribution that varies as a function of $x$ (the system is assumed invariant along the $y$ direction). For generality, we consider the slabs to be composed of anisotropic (potentially nonreciprocal) dielectric media. To simplify the analysis and avoid cumbersome $4\times4$ matrix manipulations, we impose a constraint that allows the electromagnetic problem to decouple into independent TE (electric field polarized along $y$) and TM (magnetic field polarized along $y$) polarizations. The most general anisotropic relative permittivity tensor consistent with this polarization separation is
\begin{equation} \label{eq30}
\bar{\bar{\varepsilon}}_r(\omega) =
\begin{pmatrix}
\varepsilon_{xx} & 0 & \varepsilon_{xz} \\
0 & \varepsilon_{yy} & 0 \\
\varepsilon_{zx}  & 0 & \varepsilon_{zz}
\end{pmatrix}.
\end{equation}
For the TE polarization, only the $\varepsilon_{yy}$ component contributes to the electromagnetic response. Consequently, to access a richer set of material degrees of freedom and nonlocal effects, we restrict our attention to the TM polarization. Denoting the nonlocal kernel by $F(x)$, with Fourier transform defined as $\tilde{F}(k_x)=\int_{-\infty}^{+\infty} F(x)e^{ik_x x}\,dx$, an electric-field source placed at $z=z_s$ produces, at the "image plane" $z=z_i$, the field distribution
\begin{equation} \label{eq31}
\tilde{E}_x(k_x, z_i)=\tilde{F}(k_x) \tilde{E}_x(k_x, z_s) \quad \Rightarrow \quad E_x(x, z_i) = \frac{1}{2\pi}\int_{-\infty}^{+\infty} F(x-x')E_x(x', z_s)\,dx'.
\end{equation}

To determine $\tilde{F}(k_x)$, we analyze the structure on a layer-by-layer basis using a scattering-matrix formulation, which ensures numerical stability \cite{Ref36, Ref37, Ref38}. The individual layer responses are then systematically combined to obtain the overall transfer function of the multilayer system. As an illustrative example, consider layer 2, which is bounded on the left by medium 1 (the exterior region) and on the right by medium 3, where medium 3 may represent either another layer or the exterior region. As shown in Fig.~\ref{fig4}A, for a given transverse wavenumber $k_x$, the field inside layer 2 consists of a forward-propagating wave characterized by longitudinal wavenumber $k_{z,2}^{+}$, with spatial dependence $e^{-ik_{z,2}^{+}z}$, and wave admittance $Y_2^{+}$, together with a backward-propagating wave characterized by longitudinal wavenumber $k_{z,2}^{-}$, with spatial dependence $e^{ik_{z,2}^{-}z}$, and wave admittance $Y_2^{-}$. The corresponding phases accumulated over a single traversal of the layer thickness $L_2$ are denoted by $\phi^{+}=k_{z,2}^{+} L_2$ and $\phi^{-}=k_{z,2}^{-} L_2$ for the forward- and backward-propagating waves, respectively. Defining the scattering matrix of layer 2 as $\mathbf{S}_2=\left( \begin{smallmatrix} R_2 & T_2' \\ T_2 & R_2' \end{smallmatrix} \right)$, where $R_2$ and $T_2$ denote the reflection and transmission coefficients for excitation from the left port, while $R_2'$ and $T_2'$ denote the corresponding coefficients for excitation from the right port, the scattering parameters can be expressed as
\begin{equation} \label{eq32}
\begin{gathered}
T_2 =
\frac{
t_{12} t_{23} e^{-i\phi_2^+}
}{
1 - r_{21} r_{23} e^{- i (\phi_2^++\phi_2^-)}
},  \\
R_2 =
r_{12}
+
\frac{
t_{12} t_{21} r_{23} e^{- i (\phi_2^++\phi_2^-)}
}{
1 - r_{21} r_{23} e^{- i (\phi_2^++\phi_2^-)}
}, \\
T'_2 =
\frac{
t_{21} t_{32} e^{- i \phi_2^-}
}{
1 - r_{21} r_{23} e^{- i (\phi_2^++\phi_2^-)}
}, \\
R'_2 =
r_{32}
+
\frac{
t_{23} t_{32} r_{21} e^{- i (\phi_2^++\phi_2^-)}
}{
1 - r_{21} r_{23} e^{- i (\phi_2^++\phi_2^-)}
},
\end{gathered}
\end{equation}
where these expressions are obtained using the Ray-Order Expansion method \cite{Ref39, Ref40, Ref41}. The Fresnel reflection and transmission coefficients are defined as \cite{Ref42}
\begin{equation} \label{eq33}
r_{pq} = \frac{Y_p^+ - Y_q^+}{Y_q^+ - Y_p^-}, \quad
t_{pq} = \frac{Y_p^+ - Y_p^-}{Y_q^+ - Y_p^-}.
\end{equation}
To calculate the overall scattering matrix $\mathbf{S}=\left( \begin{smallmatrix} R & T' \\ T & R' \end{smallmatrix} \right)$ of the multilayer structure, we initialize the procedure with $\mathbf{S}=\mathbf{S}_2$ and then progressively combine it with the scattering matrix of each subsequent layer, denoted $\mathbf{S}_n$. The updated scattering parameters are obtained according to
\begin{equation}\label{eq34}
\begin{gathered}
T_{\text{updated}} = T_n \,(1 - R' R_n)^{-1} \, T, \\
R_{\text{updated}} = R + T' \, R_n \,(1 - R' R_n)^{-1} \, T, \\
T'_{\text{updated}} = T' \, (1 - R_n R')^{-1} \, T_n', \\
R'_{\text{updated}} = R_n' + T_n \, R' \,(1 - R_n R')^{-1} \, T_n'.
\end{gathered}
\end{equation}
This composition method provides a numerically stable framework for determining the overall scattering characteristics of arbitrarily thick multilayer structures \cite{Ref36, Ref37, Ref38}. The stability arises because all terms involved contain only exponentially decaying factors, $e^{-i\phi}$, as evident from Eq.~(\ref{eq32}). In contrast, methods based on ABCD matrices often involve both very large exponentially growing and very small exponentially decaying terms whose combination yields a physically moderate field, leading to significant loss of numerical accuracy. If the distance between the source plane and the first layer is $d_1$ within the exterior medium labeled $1$ (assumed isotropic), and the distance between the last layer and the image plane is $d_N$ within the exterior medium labeled $N$ (also assumed isotropic), then the overall transfer function $\tilde{F}(k_x)$ is given by
\begin{equation}\label{eq35}
\tilde{F}(k_x)=\frac{T(k_x)e^{-i k_{z,1} d_1} e^{-i k_{z,N} d_N}}{1+R(k_x)e^{-2i k_{z,1} d_1}}.
\end{equation}
For further details regarding the derivation of the transfer function, the reader is referred to Appendix~A.

To determine whether $\tilde{F}(k_x)$ can represent the spatial Fourier transform of a one-sided nonlocal response, we invoke Titchmarsh’s theorem. Due to the opposite sign convention commonly adopted in the definition of the spatial Fourier transform relative to the temporal Fourier transform, the theorem requires that $\tilde{F}(k_x)$ be analytic in the upper half of the complex $k_x$ plane. The applicability of Titchmarsh’s theorem, however, depends on $\tilde{F}(k_x)$ being square integrable on the real axis. To verify this condition, it is necessary to examine the asymptotic behavior in the limit $k_x \to \infty$. Assuming isotropic exterior regions with relative permittivities $\varepsilon_1$ and $\varepsilon_N$, one finds that $k_{z,1}, k_{z,N} \to -ik_x$ in this limit, where the branch is chosen to ensure exponentially decaying waves away from the structure ($z \to \infty$). Substituting these asymptotic expressions into Eq.~(\ref{eq35}), and further assuming that the structure is passive and slightly lossy---so that the scattering coefficients $T(k_x)$ and $R(k_x)$ remain finite along the $k_x$ axis---shows that $\tilde{F}(k_x)$ decays exponentially as $k_x \to \infty$. This asymptotic behavior is favorable for square integrability, while the introduced loss further ensures that $\tilde{F}(k_x)$ is free of singularities along the real $k_x$ axis. Such singularities are generally associated with guided modes supported by the structure, and their displacement away from the real $k_x$ axis through loss-induced regularization is necessary for the rigorous application of Titchmarsh’s theorem. That said, care must be taken in selecting where the losses are introduced so as to keep the analysis analytically tractable. In this regard, it is most convenient to introduce losses only in the exterior regions (labeled $1$ and $N$), as this minimally complicates the mathematical treatment.

Following the above argument, we analyze the problem by assuming a weakly lossy isotropic exterior medium on the left side of the multilayer structure, with relative permittivity $\varepsilon_1 = 1 - i\delta$. The corresponding longitudinal wavenumber is $k_{z,1}=(\varepsilon_1 k_0^2-k_x^2)^{1/2}$, which determines the wave admittances through $Y_1^+=-Y_1^-=\omega\varepsilon_0\varepsilon_1/k_{z,1}$. We seek to analyze the branch cuts of $\tilde{F}(k_x)$ and determine whether they extend into the upper half of the complex $k_x$ plane. The quantities $Y_1^+$ and $Y_1^-$ enter the analysis only through the scattering coefficients of the first dielectric layer (region 2), as outlined in Eq.~(\ref{eq32}). In particular, all scattering parameters share the common denominator $1 - r_{21} r_{23} e^{- i (\phi_2^++\phi_2^-)}$, in which the only term associated with region 1 is $r_{21}$. Upon crossing the branch cut of $k_{z,1}$, the coefficient $r_{21}$ undergoes a discontinuity that cannot, in general, be canceled by any other factor in the denominator, since no other term depends on region 1. Moreover, the discontinuities arising in the numerators generally cannot compensate for that of the denominator, owing to their fundamentally different functional forms. As a result, all scattering parameters, and consequently $\tilde{F}(k_x)$, inherit the same branch points and branch-cut singularities as $k_{z,1}$.

Interestingly, this simple argument is sufficient to definitively determine whether one-sided nonlocality can arise in multilayer planar structures. We note that
\begin{equation} \label{eq37}
k_{z,1}=\sqrt{\varepsilon_1 k_0^2-k_x^2} \quad \Rightarrow \quad \text{Branch points}: k_x=\pm k_0\sqrt{\varepsilon_1}=\pm k_0\sqrt{1-i\delta},
\end{equation}
which, according to the reasoning above, implies that $\tilde{F}(k_x)$ possesses branch points, and consequently branch cuts, in both the upper and lower halves of the complex $k_x$ plane. As illustrated in the numerical example of Fig.~\ref{fig4}B (for a nonreciprocal gyrotropic slab), the poles likewise drift away from the real axis in both upward and downward directions upon the introduction of loss. Although establishing this analytically is considerably more involved, the existence of branch-point and branch-cut singularities in both half-planes is already sufficient to conclude that the conditions of Titchmarsh’s theorem cannot be satisfied. Therefore, a one-sided nonlocal response, namely, a one-sided spatial impulse response, cannot be realized within the framework of nonlocal flat optics based on multilayer structures (even if reciprocity is broken). We conjecture that the same conclusion applies more generally to any planar structure, including metasurfaces and photonic-crystal slabs, since they should all inherit the same branch-point and branch-cut singularities associated with open exterior regions. 

\section{Conclusion}
In this work, we investigated the foundations of causality and reviewed how they have inspired the pursuit of one-sided system responses in the temporal and spatial Fourier domains. We revisited Titchmarsh’s theorem and examined its implications for scattering problems involving spectrally causal structures and their spatial counterparts. We then turned our attention to the only remaining quadrant in the taxonomy of one-sided system responses shown in Fig.~\ref{fig1}, namely one-sided spatial nonlocality. To this end, we investigated the framework of nonlocal flat optics and demonstrated that it is fundamentally incompatible with the conditions imposed by Titchmarsh’s theorem, and therefore incapable of supporting a one-sided spatial response.

This conclusion naturally raises an intriguing question that warrants further investigation. If the framework of nonlocal flat optics fails to realize a one-sided spatial response, do alternative physical platforms exist that are capable of achieving such behavior? Or is one-sided spatial nonlocality fundamentally forbidden altogether? Furthermore, even if an exact one-sided response is unattainable within certain physical platforms, to what extent can it be approximated through optimization techniques, as done in other contexts \cite{Ref45,Ref46}? We believe that these questions are both important and intellectually compelling, potentially opening new directions for research into nonlocality and one-sided system responses. 

\section*{\label{A1}Appendix A: Analysis of Multilayer Slab Structure}

For TM fields, the electromagnetic fields can be expressed as
\begin{equation} \label{eqa1}
\mathbf{E} = (\hat{\mathbf{x}} E_x+\hat{\mathbf{z}} E_z)e^{-i(k_xx+k_zz)},
\qquad
\mathbf{H}=\hat{\mathbf{y}}H_y e^{-i(k_xx+k_zz)}.
\end{equation}
Applying Faraday’s law, $\nabla\times \mathbf{E}=-i\omega\mu_0\mathbf{H}$, yields
\begin{equation} \label{eqa2}
k_xE_z-k_zE_x=-\omega\mu_0H_y.
\end{equation}
The Ampère--Maxwell equation, $\nabla\times \mathbf{H}=
i\omega\varepsilon_0
\bar{\bar{\varepsilon}}_r\mathbf{E}$, then gives
\begin{equation} \label{eqa3}
\begin{gathered}
k_zH_y
=
\omega\varepsilon_0
(\varepsilon_{xx}E_x+\varepsilon_{xz}E_z),
\\
-k_xH_y
=
\omega\varepsilon_0
(\varepsilon_{zx}E_x+\varepsilon_{zz}E_z).
\end{gathered}
\end{equation}
The system of equations in Eq.~(\ref{eqa3}) can be solved for $E_x$ and $E_z$, yielding
\begin{equation} \label{eqa4}
\begin{pmatrix}
E_x \\ E_z
\end{pmatrix}
=
\frac{1}{\omega\varepsilon_0(\varepsilon_{xx}\varepsilon_{zz}
-
\varepsilon_{xz}\varepsilon_{zx})}
\begin{pmatrix}
\varepsilon_{zz} & -\varepsilon_{xz} \\
-\varepsilon_{zx} & \varepsilon_{xx}
\end{pmatrix}
\begin{pmatrix}
k_z \\ -k_x
\end{pmatrix}
H_y.
\end{equation}
Substituting this result into Eq.~(\ref{eqa2}) leads to the dispersion relation
\begin{equation} \label{eqa5}
\varepsilon_{xx}k_x^2
+
(\varepsilon_{xz}+\varepsilon_{zx})k_xk_z
+
\varepsilon_{zz}k_z^2
=
k_0^2
\left(
\varepsilon_{xx}\varepsilon_{zz}
-
\varepsilon_{xz}\varepsilon_{zx}
\right)
\end{equation}
which can be solved for $k_z$ for a given value of $k_x$. The corresponding wave admittance is then obtained from Eq.~(\ref{eqa4}) as
\begin{equation} \label{eqa6}
Y=\frac{H_y}{E_x}
=
\frac{\omega\varepsilon_0\left(
\varepsilon_{xx}\varepsilon_{zz}
-
\varepsilon_{xz}\varepsilon_{zx}
\right)
}{
\varepsilon_{zz}k_z
+
\varepsilon_{xz}k_x
}.
\end{equation}

Having determined the wave characteristics in each layer, the scattering parameters of the structure can be obtained using Eqs.~(\ref{eq32})--(\ref{eq34}). To derive the transfer function $\tilde{F}(k_x)$ defined in Eq.~(\ref{eq31}), we begin with
\begin{equation} \label{eqa7}
\begin{pmatrix}
a_1^{-} \\
a_N^{+}
\end{pmatrix}=
\begin{pmatrix}
R & T' \\
T & R'
\end{pmatrix}
\begin{pmatrix}
a_1^{+} \\
a_N^{-}
\end{pmatrix},
\end{equation}
where $a_1^{+}$ and $a_1^{-}$ denote the forward- and backward-propagating wave amplitudes in region 1, while $a_N^{+}$ and $a_N^{-}$ are the corresponding amplitudes in region $N$. Since $a_N^{-}=0$, the above relation reduces to
\begin{equation}\label{eqa8}
a_1^{-}= R(k_x) a_1^{+}, \qquad a_N^{+}= T(k_x) a_1^{+}.
\end{equation}
Combining these expressions with the boundary conditions
\begin{equation} \label{eqa9}
\begin{gathered}
\tilde{E}_x(k_x,z_s)=a_1^{+} e^{i k_{z,1} d_1} + a_1^{-} e^{-i k_{z,1} d_1}, \\
\tilde{E}_x(k_x,z_i)=a_N^{+} e^{-i k_{z,N} d_N},
\end{gathered}
\end{equation}
we obtain
\begin{equation} \label{eqa10}
F(k_x)=\frac{\tilde{E}_x(k_x,z_i)}{\tilde{E}_x(k_x,z_s)}
=\frac{T(k_x)e^{-i k_{z,1} d_1} e^{-i k_{z,N} d_N}}
{1+R(k_x)e^{-2i k_{z,1} d_1}}.
\end{equation}

\begin{backmatter}
\bmsection{Funding}
National Science Foundation under Award No. 2522004.

\bmsection{Disclosures}
The authors declare no conflicts of interest.

\bmsection{Data Availability Statement}
Data underlying the results presented in this paper are not publicly available at this time but may be obtained from the authors upon reasonable request.

\end{backmatter}


\bibliography{sample}






\end{document}